\documentclass[english,keywords,aps,amsmath,amssymb,twocolumn]{revtex4-1}
\usepackage[T1]{fontenc}
\usepackage[latin1]{inputenc}
\usepackage{babel}
\usepackage{amssymb}
\usepackage{graphicx}
\usepackage{color}
\usepackage{bm}
\usepackage{longtable}
\usepackage{amsmath} 
\usepackage{amsfonts}
\usepackage{amssymb}
\begin{document}
\author{A. K. Pan \footnote{akp@nitp.ac.in}}
\author{Md. Qutubuddin}
\author{Swati Kumari}
\affiliation{National Institute Technology Patna, Ashok Rajpath, Patna, Bihar 800005, India}
\title{Quantum violation of variants of Leggett-Garg Inequalities upto algebraic maximum for qubit system}

\begin{abstract} 
In 1985, Leggett and Garg formulated a class of inequalities for testing the compatibility between macrorealism and quantum mechanics. In this paper, we point out that based on the same assumptions of macrorealism that are used to derive Leggett-Garg inequalities (LGIs) , there is a scope of formulating another class of inequalities  different from standard LGIs.   By considering the three-time measurement scenario in a dichotomic system, we first propose an interesting variant of standard LGIs and show that its quantum violation is larger than the standard LGI. By extending this formulation to $n$-time  measurement  scenario, we found that the quantum violations of variants of LGIs for a \emph{qubit} system increase with $n$, and for a sufficiently large $n$ algebraic maximum can be reached. We then compare the violations of standard and variants of LGIs in unsharp measurement scenario and show that for any arbitrary $n$, the violation of later is more robust to unsharpness than the former. Further, we examine the relation between the quantum violations of the variants of LGIs and an another formulation macrorealism, known as, no-signaling in time conditions.   
\end{abstract}
\maketitle
\section{Introduction} Since the inception of quantum mechanics (QM), it remains a debatable question how our everyday world view of macrorealism can be reconciled with the quantum formalism. Historically, this question was first pointed out by Schr$\ddot{o}$dinger \cite{sch} through his famous cat  thought experiment. Since then, quite a number of attempts have been made to pose the appropriate questions relevant to this issue and to answer that questions. One effective approach to encounter this issue is to experimentally realize the quantum coherence of Schr$\ddot{o}$dinger cat-like states of large objects \cite{arndt}. Another approach within the formalism of QM is the decoherence program \cite{zur}. It explains how interaction between quantum systems and environment leads to classical behavior, but does not by itself provide the desired `cut' (\emph{\`a la} Heisenberg \cite{He25}). It is also argued that even if the decoherence effect is made negligible, the quantum behavior can be disappeared by the effect of coarse-graining of measurements\cite{bruk}.  Proposal has also been put forwarded \cite{ghi} to  modify  the dynamics of standard formalism of QM  allowing an unified description of microscopic and macroscopic systems. 

However, the above mentioned attempts do not directly address the fundamental question whether macrorealism is, in principle, compatible with the formalism of QM. Macrorealism is a classical world view that asserts that the properties of macro-objects exist independently and irrespective of ones observation. Motivated by the Bell's theorem \cite{bell64}, in 1985, Leggett and Garg \cite{leggett85} formulated a class of inequalities based on the notions of macrorealism, which provides an elegant scheme for experimentally testing the compatibility between the macrorealism and QM. 

To be more specific, the notion of macrorealism consists of two main assumptions \cite{leggett85,leggett,A.leggett} are the following;
	
\emph{ Macrorealism per se (MRps):} If a macroscopic system has two or more macroscopically distinguishable ontic states available to it, then the system remains in one of those states at all instant of time.
	
\emph{Non-invasive measurement (NIM):} The definite ontic state of the macrosystem is determined without affecting the state itself or its possible subsequent dynamics.

It is reasonable to assume that the systems in our everyday world, in principle,  obeys the aforementioned assumptions of a macrorealistic theory. Based on these assumptions, the standard Leggett-Garg inequalities (LGIs) are derived. Such inequalities can be shown to be violated in certain circumstances, thereby implying that either or both the assumptions of MRps and NIM is not compatible with all the quantum statistics. In recent times, a flurry of theoretical studies on macrorealism and LGIs have been  reported \cite{guhne,emary12,maroney14,kofler13,budroni15,budroni14,emary,halliwell16,kofler08,saha15,swati17,pan17} and a number of experiments have been performed by using various systems \cite{lambert,goggin11,knee12,laloy10,george13,knee16}.

Let us encapsulate the simplest LG scenario. Consider that the measurement of a dichotomic  observable $\hat{M}$ is performed at three different times $t_1$, $t_2$ and $t_3$ $ (t_3 \geq t_2 \geq t_1 )$. In Heisenberg picture, this in turn implies the sequential measurement of the observables  $\hat{M}_{1},\hat{M}_{2}$ and $\hat{M}_{3}$ corresponding to $t_1$, $t_2$ and $t_3$ respectively. From the assumption of MRps and NIM, one can derive the standard LGI is given by
\begin{equation}
\label{eq1}
K_{3}=\langle \hat{M_{1}} \hat{M_{2}}\rangle + \langle \hat{M_{2}} \hat{M_{3}}\rangle - \langle \hat{M_{1}} \hat{M_{3}}\rangle \leq {1} 
\end{equation}
Here $\langle {M_{1}} {M_{2}}\rangle=\sum_{m_{1},m_{2}=\pm1} m_{1},m_{2} P(M_{1}^{m_{1}}, M_{2}^{m_{2}})$  and similarly for other temporal correlation terms.  By relabeling the measurement outcomes
of each $M_i$ as $M_i = -M_i$ with $i = 1, 2,$ and $3$, three more standard LGIs can be obtained.

Instead of three times, if the measurement of $M$ is performed $n$ times, then the standard LGI for the $n$-measurement LG strings can be written as
\begin{eqnarray}
 \label{eq2}
K_{n}=\langle \hat{M}_{1} \hat{M}_{2}\rangle +...+\langle \hat{M}_{n-1} \hat{M}_{n}\rangle-\langle \hat{M}_{1} \hat{M}_{n}\rangle
\end{eqnarray}
The inequality (\ref{eq2}) is bounded as follows\cite{emary}. If $n$ is odd, $-n\leq K_{n}\leq n-2$ for $n$ $\geq{3}$  and if $n$ is even, $-(n-2)\leq K_{n}\leq n-2$ for $n$ $\geq{4}$. For $n=3$, one simply recovers inequality (\ref{eq1}).

For a two-level system, the maximum quantum value of $K_{n}$ is $(K_{n})_{Q}^{max}=n\cos\frac{\pi}{n}$. For $n=3$, $(K_{3})_{Q}^{max}=3/2$. Thus for a three-time standard LG scenario involving a dichotomic observable, the temporal Tsirelson bound of $K_3$ is $3/2$. It is proved \cite{ guhne} that this bound is irrespective of the system size.

Within the standard framework of QM, the maximum violation of CHSH inequality \cite{bell64} is restricted by the Tsirelson bound\cite{chsh}, which is significantly less than the algebraic maximum of the inequality. The  algebraic maximum may be achieved in post-quantum theory but not in QM. LGIs are often considered to be the temporal analog of Bell's inequality. However, it has been shown \cite{budroni14} that for a degenerate dichotomic observables in a qutrit system, the quantum value of $K_3$ goes up to $2.21$ and can even reach to algebraic maximum $3$ in the asymptotic limit of the system size. Such amount of violation is achieved by invoking a degeneracy breaking projective measurement which they termed as von Neumann rule. Recently, two of us have argued \cite{kumari} that such a violation of temporal Tsirelson bound has no relevance to the usual violation of LGIs.

In this paper, we first argue that by keeping the assumptions of macrorealism intact, there is scope for formulating inequalities different from the standard LGIs. We note here an important observation that due to the sequential nature of the measurement, the LG scenario is flexible than CHSH one. Such flexibility allows us to formulate new variants of standard LGIs.  For the simplest case of three-time measurement scenario, we first formulate an interesting variant of LGI and show that our proposed inequality provides considerably larger quantum violation compared to the standard LGIs. We then formulate more variants of standard LGIs by increasing number of measurements $n$ and show that the quantum violation increases with $n$. For sufficiently large $n$, the quantum values of variants of LGIs reach its algebraic maximum, even for qubit system. We note here that the calculation of $n$-time sequential correlation is a difficult task. A simplified and general formula is derived for calculating the $n$-time sequential correlation of dichotomic measurements for any arbitrary $n$. 

Further, we compare the quantum violations of the standard and variants of LGIs in unsharp measurement scenario and argue that the later is more robust to unsharpness than the former. This is due to the fact that for any arbitrary $n$ (with $n\geq3$), the  variants of LGIs can be shown to be violated for a lower value unsharpness parameter than to that of standard LGIs. Finally, we examine the relation between the variants of LGIs and another formulation of macrorealism, known as, no-signaling in time conditions.  

This paper is organized as follows. In Sec.II, we propose variant of LGI for three-time measurement scenario and demonstrate that it provide larger quantum violation compared to standard LGI. By increasing the number of measurements ($n$), in Sec.III, we formulate two more variants of LGIs. We show that for a qubit system, the quantum violation of our variants of LGIs increase with $n$ and can even reach algebraic maximum for large $n$ limit. In Sec.IV, we studied the quantum violation of variants of LGIs in the unsharp measurement scenario to show that they are more robust to unsharpness of the measurement than standard LGIs. In Sec.V, we compare variant of LGIs with standard LGIs and no-signaling in time conditions. We summarize and discuss our results in Sec.VI.

\section{Variants of LGIs in three-time measurement scenario}
We start by noting again that the standard LGIs is a particular class of inequalities but is \emph{not }unique one. The flexibility of LG scenario allows us to formulate variants of LGIs, different from the standard LGI given by Eq. (\ref{eq1}). Let us again consider the three-time LG scenario involving measurement of dichotomic  observables  $\hat{M}_{1},\hat{M}_{2}$ and $\hat{M}_{3}$ in sequence. Now, instead of three two-time correlation functions used in Eq.(\ref{eq1}), we consider a three-time correlation function $\langle \hat{M}_{1} \hat{M}_{2}\hat{M}_{3}\rangle$, a two-time function $\langle \hat{M}_{i}\hat{M}_{j}\rangle$ and finally $\langle\hat{M}_{k}\rangle$. Using them, we propose an inequality is given by
\begin{equation}
\label{3time}
K^3_{3}=\langle \hat{M}_{1} \hat{M}_{2} \hat{M}_{3}\rangle + \langle \hat{M}_{i} \hat{M}_{j} \rangle - \langle \hat{M}_{k} \rangle \leq {1}
\end{equation}
where $i,j,k=1,2,3$ with $j>i$. We call those inequalities as variant of LGIs. 

The inequalities (\ref{3time}) are violated by QM. In order to showing this, we arbitrarily choose one inequality  by taking $i,j$ and $k$ are $1,2$ and $3$ respectively, and consider the qubit state is given by 
\begin{equation}
\label{qubit}
|{\psi(t_1)}\rangle = cos \theta |{0}\rangle +  \exp(-i\phi) sin \theta |{1}\rangle
\end{equation} 
with $\theta \in [0,\pi]$ and $\phi \in [0,2\pi]$. The measurement observable at initial time $t_{1}$ is taken to be Pauli observable  $\hat{\sigma_{z}}$. The unitary evolution is given by  $U_{ij} = \exp^{-i \omega (t_{j}-t_{i})\sigma_x}$ and $\omega$ is coupling constant. For simplicity, we consider $\tau=|t_{i+1}-t_{i}|$ and  $g= \omega \tau$.

The quantum mechanical expression of $K^3_{3}$ is given by
\begin{eqnarray}
\label{eq10}
(K^3_{3})_{Q}&=& \cos2 g ( 1 + 4 \sin^{2} g \cos 2 \theta ) + 2 \sin^{2} g \cos 2\theta \nonumber\\
&-& \sin 4 g \sin 2 \theta \sin\phi
\end{eqnarray}
which is state-dependent. This is in contrast to the quantum value of standard LGI is given by 
\begin{equation}
 \label{eq9}
 (K_{3})_{Q}=2\cos 2 g - \cos 4 g
\end{equation}  
which is independent of the state.

\begin{figure}[ht]
\begin{minipage}[c]{0.55\textwidth}
\includegraphics[width=1\textwidth]{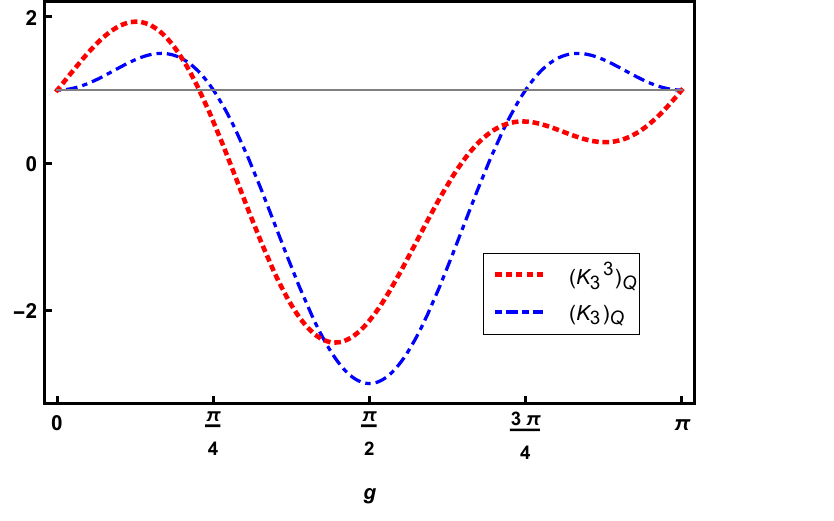}
\end{minipage}\hfill
\begin{minipage}[hc]{0.45\textwidth}
\caption{(Color online) The quantities $(K_{3})_{Q}$ and $(K_{3}^{3})_{Q}$  given by Eq.(\ref{eq10}) and Eq.(\ref{eq9}) respectively are plotted against $g$. The values of relevant parameters are $\theta= 2.66 $ and $\phi=\pi/2$. } 
\label{fig:01}
\end{minipage}
\end{figure}

If the values of the relevant parameters  are taken as $g = 0.41 $, $\theta= 2.66$ and $\phi=\pi/2$, the quantum value of $K^3_{3}$ is $1.93$, thereby violating the inequality (\ref{3time}). The maximum quantum value $(K^3_{3})$ can be shown to be $2$ for different coupling constants in between the evolutions. For simplicity, here we take same coupling constant $g$. The quantum value of $K^3_{3}$ can  then be larger than $(K_3)_{Q}^{max}=3/2$. The  quantities $(K_{3})_{Q}$ and $(K^3_{3})_{Q}$ are plotted in Figure (\ref{fig:01}).
Thus, if the larger violation of an inequality is considered to be an indicator of more non-classicality, then the variant of LGI captures the notion of macrorealism better than the standard LGIs. However, the exact correspondence between the degree of violation of LGIs and amont of non-macrorealism is not yet fully understood which requires a detailed study.
 
\section{Variants of LGIs in $n$-time measurements} 
 The above idea can be extended to  $n$-time  measurement scenario where $n>3$. For $n=4$, we propose a variant of LGI is given by
 \begin{equation}
 \label{4timek}
 K^{3}_{4}=\langle \hat{M}_{1} \hat{M}_{2} \hat{M}_{3}\hat{M}_{4}\rangle + \langle \hat{M}_{1}\hat{M}_{2} \hat{M}_{3} \rangle - \langle \hat{M}_{4} \rangle\leq {1}
 \end{equation}
which belongs to the same class of Eq. (\ref{3time}). Three more inequalities of this class can be proposed by changing the positions of $\hat{M}_{1},\hat{M}_{2}, \hat{M}_{3}$ and $\hat{M}_{4} $.

Interestingly, for $n=4$, another variant of LGI can be proposed as
\begin{equation}
\label{4timel}
 \hat{L}^{3}_{4}=\langle \hat{M}_{1} \hat{M}_{2} \hat{M}_{3} \rangle + \langle \hat{M}_{2}\hat{M}_{3} \hat{M}_{4}\rangle - \langle \hat{M}_{1}\hat{M}_{4} \rangle\leq {1}
\end{equation}
Similar to the earlier case three more inequalities can be obtained.  If number of measurements is further increased, one finds more variants of LGIs. 

Let us examine the quantum violation of inequalities (\ref{4timek}) and (\ref{4timel}) for the state given by Eq. (\ref{qubit}). The quantum mechanical expressions of $K^{3}_{4}$ and $L^{3}_{4}$ are respectively given by
\begin{eqnarray}
\label{4timekq}
(K^3_{4})_{Q}&=& \frac{1}{2} \big( 1 + \cos4 g + 8 \cos 2 g \sin ^{2} 2 g \cos2 \theta \nonumber\\
&-& 2 \sin 6 g \sin \theta\sin\phi \big)
\end{eqnarray}
\begin{eqnarray}
\label{4timelq}
(L^3_{4})_{Q}&=& 2 \cos^{2} g \cos 2 g \cos 2\theta - \cos 6 g \nonumber\\
&+& \frac{1}{2}\sin 4 g \sin 2\theta \sin\phi
\label{34a}
\end{eqnarray}
The value of $(K^3_{4})_{Q}$ is $2.12$ at $g = 1.24, \theta = 1.90$ and $\phi = \pi/2$ and of $(L^3_{4})_{Q}$ is $2.03$ at $g = 0.42, \theta = 0.21$,  and $\phi= \pi/2$. However, the above values of $(K^3_{4})_{Q}$ and $(L^3_{4})_{Q}$ are not temporal Tsirelson bound of (\ref{4timek}) and $(\ref{4timel})$, which is not very important to our present purpose. Note that, for a qubit system, the maximum quantum value of standard four-time LGI is $2\sqrt{2}$ and its macrorealist bound is 2. Then, in four-time measurement scenario, the difference between quantum and macrorealist values is $0.82$. But, in the case of our variant of LGIs, we have $(K^3_{4})_{Q}- K^3_{4}=1.12$ and $(L^3_{4})_{Q}- L^3_{4}=1.03$. It can also be seen that $(K^3_{4})_{Q}>(K^3_{3})_{Q}>(K_3)_{Q}$ and $(L^3_{4})_{Q}>(K_{3})_{Q}$. Thus, by increasing the number of measurements the quantum violation of the variants of LGIs can be improved compared to the quantum violation of standard three or four-time LGIs. 

We can then generalize the above formulation for $n$-time measurement scenario by proposing the following two inequalities are given by 
 \begin{eqnarray}
\label{ntimek}
 K^3_{n}=\langle \hat{M}_{1}\hat{M}_{2}...\hat{M}_{n}\rangle + \langle \hat{M}_{1}\hat{M}_{2}...\hat{M}_{n-1}\rangle- \langle  \hat{M_{n}}\rangle \leq{1}
\end{eqnarray}
 and
 \begin{eqnarray}
\label{ntimel}
 \hat{L}^{3}_{n}&=&\langle \hat{M}_{1}\hat{M}_{2}\hat{M}_{3}\hat{M}_{4}...\hat{M}_{n-1}\rangle + \langle \hat{M}_{2}\hat{M}_{3}...\hat{M}_{n}\rangle\nonumber\\ &-& \langle \hat{M}_{1} \hat{M}_{n}\rangle \leq{1}
\end{eqnarray}
where  $\langle \hat{M_{1}}...\hat{M_{n}}\rangle =\sum_{m_{1},...,m_{n}} m_{1}...m_{n} P(M_{1}^{m_{1}},..., M_{n}^{m_{n}})$ and similarly for other correlations.  While inequality (\ref{ntimek}) belongs to the class of (\ref{3time}) and (\ref{4timek}), the inequality (\ref{ntimel}) belongs to the other class of inequality given by (\ref{4timel}). But, both the $n$-time inequalities are derived from the same assumptions of macrorealism.

Next, we demonstrate that when $n$ is sufficiently large, the quantum values of $(K^3_{n})_{Q}$ and $(L^3_{n})_{Q}$ reach algebraic maximum of $K^3_{n}$ and $L^3_{n}$ respectively. For $n$-time sequential measurement, the calculation of correlation function in QM  is a difficult task. In order to tackle this problem, we derive a compact formula for $n$-time sequential correlation given in Eq.(\ref{s8}) of Appendix A. 

For the qubit state given by (\ref{qubit}), using Eq.(\ref{s8}) the quantum expression of $K^3_{n}$ for even $n$ is given by
\begin{eqnarray}
\label{eq11}
(K^{3}_{n_{even}})_{Q}&=& (\cos2g)^{\frac{n}{2}} + (\cos2g)^{\frac{n}{2} -1}\cos2\theta - \big( \cos 2 (n-1)g\nonumber\\ & & \cos 2\theta + \sin 2 (n-1)g \sin 2\theta \sin\phi \big)
\end{eqnarray}
and for odd $n$ 
\begin{eqnarray}
\label{eq14}
(K^{3}_{n_{odd}})_{Q}&=& (\cos 2g)^{\frac{n-1}{2}} \cos 2\theta + (\cos 2g)^{\frac{n-1}{2}} - \big( \cos 2 (n-1)g\nonumber\\ & & \cos 2\theta + \sin 2 (n-1)g \sin 2\theta \sin\phi \big)
\end{eqnarray}
By considering $g = \frac{\pi}{2n}$, Eqs.(\ref{eq11}) and (\ref{eq14}) take the form 
\begin{eqnarray}
\label{eq12}
(K^{3}_{n_{even}})_{Q}
&=& \big(\cos \frac{\pi}{n}\big)^{\frac{n}{2}} + \big(\cos \frac{\pi}{n} \big)^{\frac{n}{2}-1}\cos2\theta \nonumber\\
&+&  \cos \frac{\pi}{n} \cos2\theta- \sin \frac{\pi}{n} \sin2\theta \sin\phi 
\end{eqnarray}
and
\begin{eqnarray}
\label{eq15}
(K^{3}_{n_{odd}})_{Q}
&=& \big(\cos \frac{\pi}{n}\big)^{\frac{n-1}{2}} \cos2\theta + \big(\cos \frac{\pi}{n}\big)^{\frac{n-1}{2}} \nonumber\\
&+& \cos \frac{\pi}{n} \cos2\theta- \sin \frac{\pi}{n}\sin2\theta\sin\phi
\end{eqnarray}
respectively. In the large $n$ limit, both of them reduces to 
\begin{eqnarray}
\label{eq13}
(K^{3}_{n_{even}})_{Q}=(K^{3}_{n_{odd}})_{Q}&\approx&1+2\cos2\theta
\end{eqnarray}
Thus, when $\theta\approx 0$, the quantities $(K^{3}_{n_{even}})_{Q} = (K^{3}_{n_{odd}})_{Q}\approx 3$, i.e., the algebraic maximum of the inequalities (\ref{ntimek}-\ref{ntimel}).

We now calculate the maximum quantum value of the other variant of LGI given by (\ref{ntimel}) for the state in Eq.(\ref{qubit}). The quantum expression of  $L^{3}_{n}$ for even $n$ is given by
\begin{eqnarray}
\label{eq16}
(L^{3}_{n_{even}})_{Q}&=& (\cos 2 g)^{\frac{n}{2} - 1} \cos 2\theta + (\cos 2 g)^{\frac{n}{2} - 1} \big(\cos 2 g \cos 2\theta \nonumber\\ 
&+& \sin 2 g \sin 2\theta \sin\phi \big) - \cos 2 (n-1)g
\end{eqnarray}
and for odd $n$, we have 
\begin{eqnarray}
\label{eq18}
(L^{3}_{n_{odd})_{Q}}&=& (\cos 2g)^{\frac{n-1}{2}} + (\cos 2g)^{\frac{n-1}{2}} \nonumber\\
&-&\cos 2 (n-1)g
\end{eqnarray}
which is independent of the state.
Similar to the earlier case, by taking $g = \frac{\pi}{2n}$, from Eqs. (\ref{eq16}) and (\ref{eq18}), we have 
\begin{eqnarray}
\label{eq17}
(L^{3}_{n_{even}})_{Q}
&=& \big(\cos \frac{\pi}{n} \big)^{\frac{n}{2} - 1}\cos 2\theta  + \cos\frac{\pi}{n}+\big(\cos \frac{\pi}{n} \big)^{(\frac{n}{2} - 1)}
\nonumber\\
&& \Big( \cos \frac{\pi}{n}\cos2\theta+\sin\frac{\pi}{n} \sin2\theta \sin\phi \Big)
\end{eqnarray}
and 
\begin{eqnarray}
\label{eq19}
(L^{3}_{n_{odd}})_{Q}&=& 2 \big(\cos \frac{\pi}{n} \big)^{\frac{n-1}{2}}
+\cos\frac{\pi}{n}
\end{eqnarray}

\begin{figure}[ht]
\begin{minipage}[c]{0.5\textwidth}
\includegraphics[width=1\textwidth]{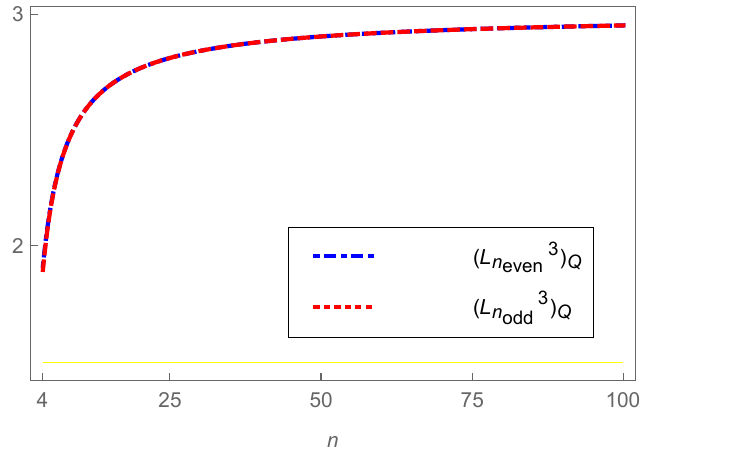}
\end{minipage}\hfill
\begin{minipage}[hc]{0.43\textwidth}
\caption{(Color online) The quantities $(L^{3}_{n_{odd}})_{Q}$ and $(L^{3}_{n_{even}})_{Q}$ given by Eqs.(\ref{eq17}) and (\ref{eq19}) respectively are plotted against number of measurements $n$ by taking $\theta=0$. Both the quantities approach algebraic maximum $3$ of the inequalities (\ref{ntimek}-\ref{ntimel})for large $n$.}
\label{fig2}
\end{minipage}
\end{figure}
For  large $n$, the quantum value of $(L^{3}_{n_{odd}})_{Q}$ approaches to algebraic maximum 3 which is independent of the state and $(L^{3}_{n_{even}})_{Q} $ approaches $3$ when  $\theta\approx 0$.  The Eqs.(\ref{eq17})  and (\ref{eq19}) are plotted in Figure \ref{fig2} to demonstrate how the quantum values of $(L^{3}_{n_{odd}})_{Q}$ and $(L^{3}_{n_{even}})_{Q}$ approach to algebraic maximum with increasing the number of measurements $n$. 

\section{Comparing standard and variants of LGIs for Unsharp Measurement scenario}
In the preceding discussion, the relevant measurements are considered to be sharp. Now, if the measurements are taken to be unsharp, it then seems interesting to examine what effect the unsharpness of the measurements would have on the quantum violations of the variants of LGIs to that for standard LGIs. Our purpose is to compare the robustness of the quantum violations of standard and variants of LGIs to the unsharpness. In order to address this issue, let us consider the sequential measurements of the unbiased POVMs  of the form
\begin{equation}
\label{GO}
M_{i}^{\pm}(0,\vec{m_{i}})=\frac{{\mathbb I} \pm \vec{m_{i}}.\sigma}{2}
\end{equation}
 At time $t_1$, we consider the POVMs as $M_{1}^{\pm}(0,\vec{m_1})$ with $\vec{m_1}=\lambda\vec{z}$, where $\lambda$ is the unsharpness parameter, with $0<\lambda\leq1$.
The time evolution of $M_{1}^{\pm}(0,\vec{m_1})$ in two different times $t_2$ and $t_3$ are given by $M_{2}^{\pm}(0,\vec{m_2})=U_{\Delta t}^{\dagger} M_{1}^{\pm}(0,\vec{m_1}) U_{\Delta t} $  and $M_{3}^{\pm}(0,\vec{m_3})=U_{2\Delta t}^{\dagger} M_{1}^{\pm}(0,\vec{m_1}) U_{2\Delta t} $ respectively. The quantum mechanical expression of  $K_ {3} $ and $K_ {3} ^ {3} $ are given by
\begin{eqnarray}
(K_{3})_{Q}&=&\lambda^2 (2 \cos 2 g - \cos 4 g )
\end{eqnarray}
\begin{eqnarray}
(K_{3}^{3})_{Q}&=&\lambda \big (\lambda  \cos 2 g (\lambda \cos 2 \theta +1)-\sin 4 g \sin 2 \theta  \sin \phi \nonumber\\
&-&\cos 4 g \cos 2 \theta \big)
\end{eqnarray}
In Figure 3, the quantities $ (K_ {3} )_{Q} $ and $ (K_ {3} ^ {3})_{Q} $ are plotted against the unsharpness parameter $\lambda$. For plotting $ (K_ {3} ^ {3})_{Q} $ the values of the relevant parameters are $\theta=2.5$, $\phi=\pi/2$ and $g=\pi/8$ and for $ (K_ {3} )_{Q} $ we take $g=\pi/6$. It is seen that from Figure 3 that the $ K_ {3} ^ {3} $ is violated for a range of unsharpness parameter 
\begin{figure}[h]
\begin{minipage}[c]{0.4\textwidth}
\includegraphics[width=\textwidth]{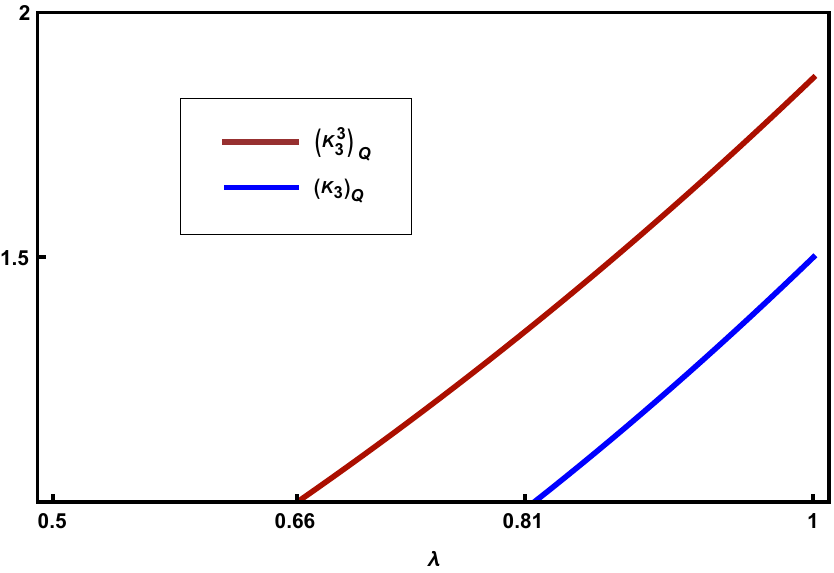}
\end{minipage}\hfill
\begin{minipage}[h]{0.45\textwidth}
\caption {The quantity  $ (K^3_ {3}) _ {Q} $ is plotted with respect to $\lambda$ by taking $\theta=2.5$, $\phi=\pi/2$ and $g=\pi/8$ and the quantity  $ (K_ {3}) _ {Q} $ is plotted by taking $g=\pi/6$. There exists a range of $\lambda \in (0.66,0.81) $, where $ K_ {3} ^ {3}$ is violated, but $K_ {3} $ does not.}
\label{fig:02-02}
\end{minipage}
\end{figure} 
 $\lambda\in (0.66,0.81) $, where $ K_ {3} $ does not. Thus, violation of variants of LGIs for three-time LG scenario is more robust than standard LGIs.

In this connection, we would also like to note that, in Ref.\cite{pan17}, Wigner form of three-time LGIs was shown to be more robust than the standard LGIs. This is due to the fact that Wigner form of LGIs violates in the range of $\lambda\in(0.69,0.81)$, where standard LGIs does not. We showed here that, variants of LGIs are even more robust that Wigner form of LGIs in three-time measurement scenario as the former violated when $\lambda\in (0.66,0.69) $, lower than the range of the violation of Wigner form of LGIs. 

A natural question arises is that whether above argument is valid for any arbitrary number of measurements $n$. For $n$-time unsharp measurement scenario, quantum expressions of $ (K_ {n})_ {Q} $, $ (K^ {3} _{n_ {even}}) _ {Q}$ and  $ (K^ {3}_ {n_ {odd}}) _ {Q} $  are respectively given by,
\begin{eqnarray}
\label{equ1}
(K_{n})_{Q}=\lambda^{2} n (\cos \frac{\pi}{n})
\end{eqnarray}
\begin{eqnarray}
\label{equ2}
(K^ {3} _ {n_ {even}}) _ {Q}
&=& \lambda^{n}\big(\cos \frac{\pi}{n}\big)^{\frac{n}{2}} +\lambda^{n-1} \big(\cos \frac{\pi}{n} \big)^{\frac{n}{2}-1}\cos2\theta \nonumber\\
&+&  \lambda (\cos \frac{\pi}{n} \cos2\theta- \sin \frac{\pi}{n} \sin2\theta \sin\phi) 
\end{eqnarray}
and
\begin{eqnarray}
\label{equ3}
(K^{3}_{n_{odd}})_{Q}
&=& \lambda^{n}\big(\cos \frac{\pi}{n}\big)^{\frac{n-1}{2}} \cos2\theta + \lambda^{n-1} \big(\cos \frac{\pi}{n}\big)^{\frac{n-1}{2}} \nonumber\\
&+&\lambda (\cos \frac{\pi}{n} \cos2\theta- \sin \frac{\pi}{n}\sin2\theta\sin\phi)
\end{eqnarray}
Note that, the above expressions are obtained by fixing $g=\pi/2n$ for which quantum value of $(K_{n})_{Q}$ is maximum.  
 
For showing the violation of $K_ {n} ^ {3} $ is more robust than $K_ {n}$ , we plotted the quantities $T_{n}^{3} =(K_ {n} ^ {3}) _ {Q} -(K_ {n} ^ {3}) _ {cl} $ and $T_{n}=(K_ {n}) _ {Q} -(K_{n}) _ {cl} $ with respect to unsharpness parameter $\lambda$ in Figure $4$. The relevant parameters are taken as $\theta=0$ and $\phi=\pi/2$. It is seen from Figure 4 that the violation of the inequalities $K_ {n}$ and $K^3_ {n}$ are obtained for $\lambda\in[0.81,1]$ and $\lambda\in[0.75,1]$ respectively. Thus, for the range of $\lambda\in[0.75,0.81]$, $K^3_ {n}$ is violated but no violation of $K_ {n}$ occurs. We already know that for $n=3$, the lowest value of $\lambda$ is $0.66$ which is obtained for $g=\pi/8$. In Figure 4 we take $g=\pi/2n$, i.e., for $n=3$ we have $g=\pi/6$ and for which the lowest value $\lambda$ changes to  $0.75$.

\begin{figure}[h]
\begin{minipage}[c]{0.42\textwidth}
\includegraphics[width=\textwidth]{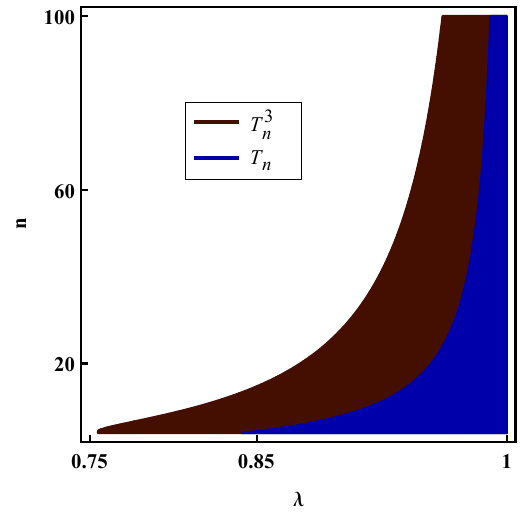}
\end{minipage}\hfill
\begin{minipage}[h]{0.45\textwidth}
\caption {(Color online) The quantity $T_{n}= (K_ {n}) _ {Q}-(K_ {n}) _ {cl} $ and  $T_{n}^{3}= (K^3_ {n}) _ {Q} -(K^3_ {n}) _ {cl} $ are plotted against unhappiness parameter ($\lambda$) for $\theta=0$ and $\phi=\pi/2$. The violation of the inequality $K^3_ {n}$ is obtained for a range of $\lambda\in[0.77,0.81]$, where no violation of $K_ {n}$ occur.}
\label{fig:02-02}
\end{minipage}
\end{figure}

 Similarly, for comparing the robustness of the violations of $L^ {3}_{n}$ and $K_{n}$ in unsharp measurement scenario, we write down the quantum expressions of $ (L^ {3}_ {n_ {even}}) _ {Q} $ and $ (L^ {3} _ {n_ {odd}) _ {Q}} $ calculated by taking $g=\pi/2n$ are respectively given by
 \begin{eqnarray}
\label{equ4}
(L^{3}_{n_{even}})_{Q}&=&\lambda^{n-1}(\cos \frac{\pi}{n})^{\frac{n}{2} - 1}\nonumber\\
&\times&\big(\cos \frac{\pi}{n}\cos 2\theta + \sin \frac{\pi}{n}\sin 2\theta\sin\phi \big)\nonumber\\  
&+&\lambda^ {n-1} (\cos \frac{\pi}{n})^{\frac{n}{2} - 1} \cos 2\theta +\lambda^{2} \cos \frac{\pi}{n} 
\end{eqnarray}
and
\begin{eqnarray}
\label{equ5}
(L^{3}_{n_{odd})_{Q}}= 2\lambda^{n-1} (\cos\frac{\pi}{n})^{\frac{n-1}{2}}
+\lambda^ {2} \cos \frac{\pi}{n}
\end{eqnarray}
We plotted the quantities $T_{n}$ and  $U^3_ {n}=(L^3_ {n}) _ {Q} -(L^3_ {n}) _ {cl} $ against $\lambda$ in Figure $5$ by taking the values of the relevant parameters as $\theta=0$ and $\phi=\pi/2$. It is seen from Figure 5 that the violation of the inequalities $K_ {n}$ and $L^3_ {n}$ are obtained for  $\lambda\in[0.77,1]$ and $\lambda\in[0.85,1]$ respectively. Thus, for a range of $\lambda\in[0.77,0.85]$, $L^3_ {n}$ is violated but no violation of $K_ {n}$ occurs, which leads us to conclude that the violation of $L_ {n} ^ {3}$ is more robust than $K_ {n} $.

\begin{figure}[ht]
\begin{minipage}[c]{0.42\textwidth}
\includegraphics[width=1\textwidth]{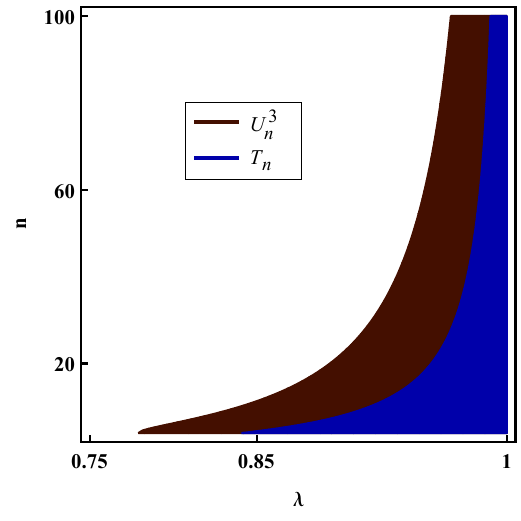}
\end{minipage}\hfill
\begin{minipage}[hc]{0.43\textwidth}
\caption {(Color online) The quantity $T_{n}=(K_ {n}) _ {Q} -(K_ {n}) _ {cl} $ and  $U^3_ {n}=(L^3_ {n}) _ {Q} -(L^3_ {n}) _ {cl} $ are plotted against unsharpness parameter ($\lambda$) for $\theta=0$ and $\phi=\pi/2$.  $L^3_ {n}) _ {Q}$ is found to be more robust than $K_ {n}) _ {Q}$ for any value of $n$.}
\label{fig2}
\end{minipage}
\end{figure} 

We note here that with increasing $n$, the critical value of $\lambda$ (say, $\lambda_{c}$) beyond which the quantum violation of variants of LGIs and standard LGIs occur increases. For a sufficiently large value of $n$, the quantum violation can only be obtained for sharp measurement $(\lambda=1)$. But, for any arbitrary $n$ the suitable parameters can be found for which $\lambda_c$ is significantly lower in the case of variants of LGIs than the standard ones. Hence, for any arbitrary $n$, the violation of variants of LGIs is more robust than the standard LGIs in the unsharp measurement scenario and a better candidate  for testing the macrorealism than the standard LGIs. 

\section{Comparing variants of LGIs with other formulations of macrorealism}
Fine \cite{fine} theorem states that the CHSH inequalities are necessary and sufficient condition for local realism. Since standard LGIs are often considered to be the temporal analogue of CHSH inequalities one may expect that they also provide the necessary and sufficient condition for macrorealism. In their recent works, Clemente and Kofler \cite{clemente} showed that no set of  standard LGIs can provide the necessary and sufficient condition for macrorealism. In this connection we would like to mention that  two of us \cite{swati17} have shown that the Wigner formulation of LGIs are stronger than standard LGIs but they also do not provide necessary and sufficient condition for macrorealism. 
 
Clemente and Kofler \cite{clemente} introduced a very different conditions for macrorealism formulated in terms of no-signalling in time (NSIT) conditions that does not involve LGIs. NSIT condition is the statistical version of NIM condition. It is analogous to the no-signaling in space condition in Bell's theorem, but violation of NSIT condition does not provide any inconsistency with physical theories. It simply assumes that the probability of an outcome of measurement remains unaffected due to prior measurement. Clearly, the satisfaction of all NSIT conditions in any operational theory ensures the existence of global joint probability condition $P(M_{1}^{m_1},M_{2}^{m_2},M_{3}^{m_3})$ where $m_1,m_2,m_3=\pm1$ and in such a case no violation of any LGI can occur. They \cite {clemente} showed that a suitable conjunction of NSIT conditions provides the necessary and sufficient conditions for macrorealism. Above discussion may the  lead one to conclude that formulation of macrorealism based on NSIT conditions is better test than LGIs.

In an interesting work Halliwell \cite{halliwell17} have recently argued that LGIs and NSIT conditions capture two different notions of macrorealism. The difference lies in the ways that are adopted for implementations of NIM conditions. Since the inception of LGIs the requirement of the satisfaction of NIM in measurement remains a source of considerable debate. The statistical version of locality assumption in a realist model, i.e., the no-signalling in space is satisfied in any operational theory including quantum theory. On the other hand, statistical version of non-invasive measurability, i.e., the NSIT condition is not, in general, satisfied in quantum theory. The experimental test of LGIs needs to guarantee how non-invasive measurability in macrorealist model can be justified. Halliwell \cite{halliwell17} discussed that standard LGIs are the test of a weaker form of macrorealism (MR$_{weak}$) which may be considered as a direct test of MRps. On the other hand, NSIT conditions test a stronger form of macrorealism (MR$_{strong}$) basically proposes a test of NIM, and in QM, it is in general violated. 

From the above perspective, it would then be interesting to study the notion of macrorealism involved in the variants of LGIs and its relation to the standard LGIs and NSIT formulation of macrorealism. This subtle conceptual issues  would be of considerable interest and will be discussed in detail elsewhere. What we would like to show in this Section is the relation between variant of LGIs and NSIT condition. In order to demonstrate this, we first re-write the standard LGIs in terms of NSIT condition. We show that the NSIT conditions are necessary for the standard LGIs but no sufficient. Mere violation of various NSIT conditions are not enough to warrant the violation of LGIs. This is in agreement with the argument presented in \cite{halliwell17}. Since the NSIT conditions correspond to a stronger notion of macrorealism, the violations of them may not guarantee the violation of weaker form of macrorealism involved in standard LGIs. Then, it is justified that no set of standard LGIs provides the necessary and sufficient conditions for for the notion of macrorealism captured by NSIT condition. Further, we analyze the variant of LGIs for three-time measurement scenario in this context.

A two-time NSIT condition can be written as 
\begin{equation}
NSIT_{(1)2}:P(M_{2}^{m_2})=\sum_{m_1} P_{12}(M^{m_1}_{1},M^{m_2}_{2})
\end{equation}
which means that the probability $P(M_{2}^{m_2})$ is unaffected by the prior measurement of $M_{1}$. Similarly, a three-time NSIT condition is given by
\begin{eqnarray}
\nonumber
NSIT_{(1)23}:P(M_{2}^{m_2},M_{3}^{m_3})&=&\sum_{m_1}P_{123}(M_{1}^{m_1},M_{2}^{m_2},M_{3}^{m_3})\\
\end{eqnarray}
Here $P_{123}(M_{1}^{m_1},M_{2}^{m_2},M_{3}^{m_3})$ denotes the joint probabilities when all the three measurements are performed.

Clemente and Kofler \cite{clemente} have shown that a suitable conjunction of two-time and three-time NSIT conditions provides the necessary and sufficient condition for macrorealism, i.e.,
\begin{eqnarray}
\label{nsit}
NSIT_{(2)3} \wedge NSIT_{(1)23} \wedge NSIT_{1(2)3} \Leftrightarrow MR_{strong}
\end{eqnarray}
where MR$_{strong}$ denotes stronger notion of macrorealism as argued by Halliwell\cite{halliwell17}. 

We first show the reason the standard LGIs do not provide necessary and sufficient condition for MR$_{strong}$. Such an argument was first initiated in \cite{maroney14} and discussed in detail in  \cite{swati17}. But for making the present work self-contained we encapsulate the essence of the argument.  

Let us consider the pairwise marginal statistics of the experimental arrangement when all three measurements ($M_1$, $M_2$ and $M_3$) are performed and introduce the following quantity 
\begin{eqnarray}
\nonumber
D_{1}(M_{2}^{m_2},M_{3}^{m_3})&=&P(M_{2}^{m_2},M_{3}^{m_3})\\
\label{d11}
&-&\sum_{m_1} P_{123}(M_{1}^{m_1},M_{2}^{m_2},M_{3}^{m_3})
\end{eqnarray}
which quantifies the amount of disturbance created (in other words, degree of violation of NSIT condition) by the measurement $M_1$ at $t_1$ to the measurements of $M_2$ and $M_3$ at $t_2$ and $t_3$ respectively. Similarly,
\begin{eqnarray}
\nonumber
D_{2}(M_{1}^{m_1},M_{3}^{m_3})&=&P(M_{1}^{m_1},M_{3}^{m_3})\\
\label{d22}
&-&\sum_{m_2} P_{123}(M_{1}^{m_1},M_{2}^{m_2},M_{3}^{m_3})
\end{eqnarray}
\begin{eqnarray}
\nonumber
D_{3}(M_{1}^{m_1},M_{2}^{m_2})&=&P(M_{1}^{m_1},M_{2}^{m_2})\\
\label{d33}
&-&\sum_{m_3} P_{123}(M_{1}^{m_1},M_{2}^{m_2},M_{3}^{m_3})
\end{eqnarray}
Note that, since no information can travel backward in time, $D_{3}(M_{1}^{m_1},M_{2}^{m_2})=0$ in any physical theory. For two-time measurements, we can define similar quantity, for example, $D_{1}(M_{2}^{m_2})$. 

Standard LGIs can be derived by assuming the satisfaction of all NSIT conditions. But, in QM, the NSIT conditions are, in general, not satisfied. It is then straightforward to understand that the difference between $K_3$ and $(K_3)_{123}$ plays an important role for the violation of LGI. Clearly, if $K_{3}=(K_{3})_{123}$ is satisfied, the LGI will \textit{not} be violated.  When all the three measurements are performed for measuring each correlation, the expression of $K_{3}$ in inequality(\ref{eq1}) can be written 
\begin{eqnarray}
\nonumber
(K_{3})_{123}&=& \langle M_{1}M_{2}\rangle_{123}
+\langle M_{2}M_{3}\rangle_{123}-\langle M_{1}M_{3}\rangle_{123}\nonumber\\
\label{d27}
&=&  1-4\alpha
\end{eqnarray}
where $\alpha =P(M_{1}^{+},M_{2}^{-},M_{3}^{+})+P(M_{1}^{-},M_{2}^{+},M_{3}^{-})$. 

Using Eqs.(\ref{d11}) and (\ref{d22}) we can write
\begin{eqnarray}
\label{xx}
&& K_{3}-(K_{3})_{123}=\\
\nonumber
&&\sum_{m_2 =m_3}D_{1}(M_{2}^{m_2},M_{3}^{m_3})-\sum_{m_1= m_3}D_{2}(M_{1}^{m_1},M_{3}^{m_3})\\
\nonumber
&-&\sum_{m_2 \neq m_3} D_{1}(M_{2}^{m_2},M_{3}^{m_3})+\sum_{m_1\neq m_3}D_{2}(M_{1}^{m_1},M_{3}^{m_3})
\end{eqnarray}
Since, $\sum D_{1}(M_{2}^{m_2},M_{3}^{m_3})=0$, and $\sum D_{2}(M_{1}^{m_1},M_{3}^{m_3})=0$ and $K_{3}\leq 1$, from Eq.(\ref{xx})we obtain
\begin{eqnarray}
\nonumber
&&2\sum_{m_2=m_3}D_{1}(M_{2}^{m_2},M_{3}^{m_3})-2\sum_{m_2=m_3}D_{2}(M_{1}^{m_1},M_{3}^{m_3})\\
&+&(K_{3})_{123}\leq 1
\end{eqnarray}
	By putting the value of $(K_{3})_{123}$ from Eq.(\ref{d27}) we have
\begin{eqnarray}
\sum_{m_2=m_3}D_{1}(M_{2}^{m_2},M_{3}^{m_3})-\sum_{m_1=m_3}D_{2}(M_{1}^{m_1},M_{3}^{m_3})\leq 2\alpha\nonumber\\
\end{eqnarray}
We have thus written down the standard LGIs in terms of NSIT conditions. For the violation of standard LGI in (\ref{eq1}) the relation 
\begin{eqnarray}
\label{lgsatis}
\sum_{m_2=m_3}D_{1}(M_{2}^{m_2},M_{3}^{m_3})-\sum_{m_1=m_3}D_{2}(M_{1}^{m_1},M_{3}^{m_3}) > 2\alpha\nonumber\\
\end{eqnarray}
needs to be satisfied in QM. This implies that for violation of standard LGI at least one of the two three-time NSIT conditions ($NSIT_{(1)23}$ and $NSIT_{1(2)3}$) required to be violated. However, mere violations of NSIT conditions do not guarantee the violation of LGIs which depends on the interplay between the violations of two NSIT conditions and on a threshold value $2\alpha$. A similar discussion\cite{halliwell17} has been given in terms of quantum witness.  Thus, NSIT conditions are necessary for LGI but not sufficient \cite{maroney14, swati17}. Then the violations of NSIT conditions provide the necessary and sufficient condition for MR$_{strong}$ but not for MR$_{weak}$ captured by LGIs. 

Next, we compare our variant of LGIs with standard LGIs and NSIT conditions.  Let us write one of the variant of LGIs for three-time measurement scenario is given by
\begin{equation}
\label{3t}
K^3_{3}=\langle \hat{M}_{1} \hat{M}_{2} \hat{M}_{3}\rangle + \langle \hat{M}_{1} \hat{M}_{2} \rangle - \langle \hat{M}_{3} \rangle \leq {1}
\end{equation}
Before writing variant of LGI in terms of NSIT conditions, we note the following interesting point.  Since $\langle \hat{M}_{1} \hat{M}_{2} \hat{M}_{3}\rangle=(\langle \hat{M}_{1} \hat{M}_{2} \hat{M}_{3}\rangle)_{123}$ and $\langle \hat{M}_{1} \hat{M}_{2} \rangle=(\langle \hat{M}_{1} \hat{M}_{2} \rangle)_{123}$, then disturbance can only come due to the term $\langle \hat{M}_{3} \rangle $. Intuitively one may then expect that whenever the quantity $D_{12}(M_{3}^{m_{3}})$ defined as 
\begin{eqnarray}
\label{d222}
D_{12}(M_{3}^{m_{3}})=P(M_{3}^{m_{3}})-\sum_{m_{1},m_{2}} P_{123}(M_{1}^{m_{1}},M_{2}^{m_{2}},M_{3}^{m_{3}})\nonumber\\
\end{eqnarray}
is nonzero, the violation of the variant of LGI given by Eq.(\ref{3t}) can be obtained. Thus, one may expect that the NSIT condition $NSIT_{(12)3}$ provides the necessary and sufficient condition for variant of LGI. But, we shall shortly see that similar to the case of standard LGI, $NSIT_{(12)3}$  provides the necessary but not the sufficient condition for the violation of variants LGIs.
 
Using  similar approach adopted for standard LGIs, we express the variant of LGI given by Eq.(\ref{3t}) in terms of NSIT condition.  Then, when all three measurements are performed the expression of $K_{3}^{3}$ in Eq.(\ref{3t}) can be written as 
\begin{eqnarray}
\nonumber
(K_{3}^{3})_{123}&=& \langle \hat{M}_{1}\hat{M}_{2}\hat{M}_{3}\rangle_{123}
+\langle \hat{M}_{1}\hat{M}_{2}\rangle_{123}-\langle \hat{M}_{3}\rangle_{123}\\
\label{beta}
&=&  1-4\beta
\end{eqnarray}
where $\beta =P(M_{1}^{+},M_{2}^{-},M_{3}^{+})+P(M_{1}^{-},M_{2}^{+},M_{3}^{+})$. Using Eq.(\ref{d222}),  we can write
\begin{eqnarray}
K_{3}^{3}-(K_{3}^{3})_{123}=D_{12}(M_{3}^{-})-D_{12}(M_{3}^{+})
\label{d12}
\end{eqnarray}
 Since $	K_{3}^{3}\leq 1$, using Eq.(\ref{beta}) we obtain
\begin{eqnarray}
	D_{12}(M_{3}^{-})\leq 2\beta
\end{eqnarray}
For the violation of $K_3^{3}$ in Eq. (\ref{3time}) the following relation needs to be satisfied in QM is given by
\begin{eqnarray}
\label{dis}
D_{12}(M_{3}^{-}) > 2\beta
\end{eqnarray}
Comparing Eq.(40) with Eq.(46), it is seen that two NSIT conditions and their interplay was involved in Eq. (40) instead of single NSIT condition in (\ref{dis}). However, mere violation of $NSIT_{(12)3}$ does not provide the violation of $K_3^{3}$, the value of $D_{12}(M_{3}^{-})$ needs to greater than a threshold value $2\beta$. Thus, NSIT condition is necessary for the violation of variant of LGI but not sufficient. In other words, the violation of NSIT condition provide the violation of MR$_{strong}$ but do not provide the violation of notion of macrorealism captured by variants of LGIs.. 

This argument can be made more interesting as follows. Let the coupling between the evolutions are different, say, for $M_1$ to $M_2$ the coupling is $g_1$ and for $M_2$ to $M_3$ it is $g_2$. For suitable choices of relevant parameters, $g_1 =\pi, \theta=0$ and $\phi=\pi/2$, the value of $\beta=0$. In such a case, from Eq.(46), we have $D_{12}(M_{3}^{-}) > 0$ is the condition of violation of variant of LGIs which is nothing but the violation of $NSIT_{(12)3}$. We can then say that in some special cases, the violation of NSIT condition provides the violation of macrorealism involved in LGIs. Such argument may not made for the case of standard LGIs. We would like to remark here that there is no clear consensus in the literature regarding the different notions of macrorealism involved in LGIs and NSIT conditions.  Detailed study along this line is needed to understand such notions macrorealism. 
\section{Summary and Discussion}
The quantum violation of standard LGIs for a dichotomic system is restricted by temporal Tsirelson bound which is significantly lower than the algebraic maximum. In this paper, we note an important observation that the standard LGIs are a class of inequalities but \emph{not} the unique one. There is a scope of formulating new variant of inequalities based on the assumptions of MRps and NIM. For the simplest case of three-time measurement scenario, we first proposed new variants of LGIs which are different from the standard LGIs. For a qubit system, we demonstrated that such macrorealist inequalities provide larger quantum violation than standard LGIs. By increasing the number of measurements $n$, we proposed more variants of LGIs. We found that  the quantum violation of variants of LGIs increase with the increment of $n$. Interestingly, for a sufficiently large value of $n$, the quantum violation of variant of LGIs reach their algebraic maximum, even for qubit system. 

In the context of unsharp measurement scenario, we compared the quantum violations of standard and variants of LGIs. It is seen that for any arbitrary number of measurements $n$, the variant of LGIs is violated for a lower value of unsharpness parameter $\lambda$ compared to the standard ones. This result enables us to conclude that the variants of LGIs is more robust and provide a better test of macrorealism in unsharp measurement scenario.

Further, we compared the variants of LGIs with another formulation of macrorealism known as NSIT conditions. NSIT condition is the statistical version of NIM assumption of macrorealism. As discussed in \cite{halliwell17} that the standard  LGIs and NSIT conditions capture two different notions of macrorealism. The difference arises due to the different ways of implementing the NIM conditions. In LGIs, the NIM condition is weaker than that is involved in NSIT condition. In other words, LGIs test a weaker form of macrorealism in contrast to NSIT conditions. LGIs can provide the necessary and sufficient condition for MR$_{weak}$ but not for MR$_{strong}$ \cite{clemente, halliwell17}. By introducing the degree of violation of NSIT condition, we first point out why the NSIT conditions are necessary but not sufficient criteria for standard LGI. We then showed that the violation of variants of LGIs require only one NSIT condition to be violated. However, the violation needs to be beyond a threshold value. In the context of the discussion initiated in \cite{halliwell17}, we remark that the variants of LGIs captures a form of macrorealism which  may differ from the notion of macorealism  involved in standard LGIs. This issue could be an interesting avenue of future research. 

Finally, we note that in the study regarding nonlocality, it is  recently argued \cite{caval}  that the violation of Bell's inequality may be a poor quantifier of the amount of nonlocality. In this paper, we have demonstrated the amount of quantum violation of variants of LGIs is larger than standard LGIs. While the violation of a LGI implies the violation of a notion of macrorealism, the exact correspondence between degree of quantum violation of LGIs and amount of non-macrorealism  has not been explored yet. This calls for further study.  
\section*{Acknowledgments}
AKP acknowledges the support from Ramanujan Fellowship research grant (SB/S2/RJN-083/2014). MQ  acknowledge the Junior Research Fellowship from SERB project (ECR/2015/00026). 

\appendix
\begin{widetext}
\section{General formula for calculating the   sequential correlation of $n$-measurement}
We provide a general formula for calculating sequential correlation of $n$-time measurements of a dichotomic observable. In LG scenario, the measurement of a dichotomic observable  $\hat{M}$ having outcomes $\pm1$ is performed at time $t_1$,  $t_2$.... $t_n$ $(t_1<t_2<...<t_n)$, which, in turn, can be considered as the sequential measurement of the observables $\hat{M}_1$, $\hat{M}_2$.... and $\hat{M}_n$ respectively. 

Given a density matrix $\rho$, the correlation function for the sequential measurement of two observables $\hat{M}_1$ and $\hat{M}_2$ can be calculated by using the formula \cite{budroni14}
\begin{eqnarray}
\langle \hat{M}_{1} \hat{M}_{2}\rangle_{seq}=\frac{1}{2}Tr\left[\rho\left\{\hat{M}_{1},\hat{M}_{2}\right\}\right]
\end{eqnarray} 
where $\left\{\right\}$ denotes anti-commutation.

We generalize the above formula for $n$-time measurement scenario. For this, let us first consider the three-measurement scenario as an example. The correlation function for three-time measurement can be written as,
\begin{eqnarray}
\label{s2}
\langle \hat{M}_{1}\hat{M}_{2}\hat{M}_{3}\rangle_{seq}=\sum_{m_{1},m_{2},m_{3}=\pm1}{m_{1}m_{2}m_{3}}P( M_{1}^{m_{1}}M_{2}^{m_{2}}M_{3}^{m_{3}})\nonumber\\
\end{eqnarray}
Let $\Pi_{M_{1}}^{m_{1}}$, $\Pi_{M_{2}}^{m_{2}}$ and $\Pi_{M_{3}}^{m_{3}}$ are projectors of observables $\hat{M}_{1}$, $\hat{M}_{2}$ and $\hat{M}_{3}$  corresponding to the to eigenvalues $m_{1},m_{2}$ and $m_{3}$ respectively. In QM, Eq.(\ref{s2}) can then be written as,
\begin{eqnarray}
\label{s3}
\langle \hat{M}_{1}\hat{M}_{2}\hat{M}_{3}\rangle_{seq}&=&\sum_{m_{1},m_{2},m_{3}=\pm1}{m_{1}m_{2}m_{3}}Tr[\Pi_{M_{2}}^{m_{2}}\Pi_{M_{1}}^{m_{1}}\rho\Pi_{M_{1}}^{m_{1}}\Pi_{M_{2}}^{m_{2}}\Pi_{M_{3}}^{m_{3}}]\nonumber\\
&=&\sum_{m_{1},m_{2}=\pm1}{m_{1}m_{2}}Tr[\Pi_{M_{2}}^{m_{2}}\Pi_{M_{1}}^{m_{1}}\rho\Pi_{M_{1}}^{m_{1}}\Pi_{M_{2}}^{m_{2}}\Pi^{+}_{M_{3}}]-\sum_{m_{1},m_{2}=\pm1}{m_{1}m_{2}}Tr[\Pi_{M_{2}}^{m_{2}}\Pi_{M_{1}}^{m_{1}}\rho\Pi_{M_{1}}^{m_{1}}\Pi_{M_{2}}^{m_{2}}\Pi^{-}_{M_{3}}]
\end{eqnarray}
Using $\hat{M}_{3} =\Pi_{M_{3}}^{+}-\Pi_{M_{3}}^{-}$ and putting the value of $m_2=\pm 1$, we have 
\begin{eqnarray}
\label{s5}
\langle \hat{M}_{1}\hat{M}_{2}\hat{M}_{3}\rangle_{seq}&=&\sum_{m_{1}=\pm1}{m_{1}}Tr[(\Pi_{M_{2}}^{+}\Pi_{M_{1}}^{m_{1}}\rho\Pi_{M_{1}}^{m_{1}}\Pi_{M_{2}}^{+}).{\hat{M}_{3}}]-\sum_{m_{1}=\pm1}{m_{1}}Tr[(\Pi_{M_{2}}^{-} \Pi_{M_{1}}^{m_{1}} \rho \Pi_{M_{1}}^{m_{1}} \Pi_{M_{2}}^{-}).{\hat{M}_{3}}] 
\end{eqnarray} 
Since $\Pi_{M_{2}}^{\pm1}=(\mathbb{I}\pm\hat{M}_{2})/2$,  Eq.(\ref{s5}) can be simplified as 
\begin{eqnarray}
\label{s6}
\langle\hat{M}_{1}\hat{M}_{2}\hat{M}_{3}\rangle_{seq}=\frac{1}{2}\sum_{m_{1}=\pm1}{m_{1}}Tr\left[(\Pi_{M_{1}}^{m_{1}}\rho\Pi_{M_{1}}^{m_{1}}).\left\{\hat{M}_{2},\hat{M}_{3}\right\}\right] 
\end{eqnarray}
Adopting the similar to the procedures adopted above, further simplification provides
\begin{eqnarray}
\label{s7}
\langle\hat{M}_{1} \hat{M}_{2}\hat{M}_{3}\rangle_{seq}=\frac{1}{4}Tr\left[\rho\left\{\hat{M}_{1},\left\{\hat{M}_{2},\hat{M}_{3}\right\}\right\}\right]
\end{eqnarray}
For the case of $n$-time measurements, we derive
\begin{eqnarray}
\label{s8}
\langle \hat{M}_{1}\hat{M}_{2}.......\hat{M}_{n-1}\hat{M}_{n}\rangle_{seq}=\frac{1}{2^{n-1}}Tr\left[\rho\left\{\hat{M}_{1},\left\{\hat{M}_{2},........,\left\{\hat{M}_{n-2},\left\{\hat{M}_{n-1},\hat{M}_{n}\right\}\right\}\right\}\right\}\right]
\end{eqnarray}
\end{widetext}
\end{document}